# Development of a Respiratory Sound Labeling Software for Training a Deep Learning-Based Respiratory Sound Analysis Model

Fu-Shun Hsu[ab], Chao-Jung Huang[c], Chen-Yi Kuo[b], Shang-Ran Huang[b], Yuan-Ren Cheng[bde], Jia-Horng Wang[f], Yi-Lin Wu[b], Tzu-Ling Tzeng[b], Feipei Lai*[a]

[a]Graduate Institute of Biomedical Electronics and Bioinformatics, National Taiwan University, Taipei, Taiwan; [b]Heroic Faith Medical Science Corporation Limited, New Taipei, Taiwan; [c]National Taiwan University-Stanford Joint Program Office of AI in Biotechnology, Ministry of Science and Technology Joint Research Center for Artificial Intelligence Technology and All Vista Healthcare, Taipei, Taiwan; [d]Department of Life Science, College of Life Science, National Taiwan University, Taipei, Taiwan; [e]Institute of Biomedical Sciences, Academia Sinica, Taipei, Taiwan; [f]Department of Critical Care, Far Eastern Memorial Hospital, New Taipei City, Taiwan
*flai@csie.ntu.edu.tw

**ABSTRACT**

Respiratory auscultation can help healthcare professionals detect abnormal respiratory conditions if adventitious lung sounds are heard. The state-of-the-art artificial intelligence technologies based on deep learning show great potential in the development of automated respiratory sound analysis. To train a deep learning-based model, a huge number of accurate labels of normal breath sounds and adventitious sounds are needed. In this paper, we demonstrate the work of developing a respiratory sound labeling software to help annotators identify and label the inhalation, exhalation, and adventitious respiratory sound more accurately and quickly. Our labeling software integrates six features from MATLAB Audio Labeler, and one commercial audio editor, RX7. As of October, 2019, we have labeled 9,765 15-second-long audio files of breathing lung sounds, and accrued 34,095 inhalation labels, 18,349 exhalation labels, 13,883 continuous adventitious sounds (CASs) labels and 15,606 discontinuous adventitious sounds (DASs) labels, which are significantly larger than previously published studies. The trained convolutional recurrent neural networks based on these labels showed good performance with F1-scores of 86.0% on inhalation event detection, 51.6% on CASs event detection and 71.4% on DASs event detection. In conclusion, our results show that our proposed respiratory sound labeling software could easily pre-define a label, perform one-click labeling, and overall facilitate the process of accurately labeling. This software helps develop deep learning-based models that require a huge amount of labeled acoustic data.

## 1. INTRODUCTION

Physicians routinely listen to lung sounds through respiratory auscultation during general examinations which could help healthcare professionals detect abnormal respiratory conditions if adventitious lung sounds are heard. Such respiratory auscultation is an important tool for physicians in making clinical decisions. Many studies have been conducted on the development of automated respiratory sound analysis based on deep learning methods. Messner et al. proposed an event detection approach with spectral features and bidirectional gated recurrent neural networks (BiGRNNs) and had around 86% F1-score on breath phase detection [1]. Jácome, et al. applied the well-known Faster R-CNN (FasterRCNN) object detection system and achieved an average sensitivity of 97% and an average specificity of 84% on breath phase detection in lung sound recordings [2]. Regarding using deep learning in adventitious sounds analysis, Chen et al. proposed an optimized S-transform and deep residual networks (ResNets) method for the recognition of wheezes, crackles and normal respiratory sounds, and the experimental results showed the outperformance of the proposed model with the classification accuracy of 98.79% [3].

Lung sounds naturally have complex structures and are non-stationary signals. This property can be observed both in normal breathing sounds and adventitious breathing sounds. Lung sounds may be classified according to two main categories: normal breathing sounds or adventitious lung sounds. Generally, normal breathing sounds that are heard when no respiratory disorders exist and adventitious sounds are heard when a respiratory disorder exists [4], [5]. Adventitious sound can be further classified into two groups, continuous adventitious sounds (CASs) and discontinuous adventitious sounds (DASs) [6], [7]. CASs have many subtypes, such as wheezes, stridor, and rhonchi. DASs majorly consist of coarse crackles, and fine crackles. If accurately identified, the presence of these sounds usually indicates an important diagnostic value.

A great amount of research has been focused on the automatic detection or classification of adventitious respiratory sound studies in the past few decades. Semedo et al. [8] introduced Computerized Lung Auscultation Sound System (CLASS), the first open sourced tool that can record respiratory sound and analyze them simultaneously. Pramono et al. [7] summarized that the most relevant algorithms developed to detect or classify events usually involve two steps: the first step is to extract the relevant features that will be used as detection or classification variables, and the second step is to use detection or classification techniques on the data, based on the features extracted.

To develop a detection or classification algorithm, the first step is to establish a dataset with ground truth labels. However, accurate labeling of collected respiratory sound is a challenging task because of varying perception of sound and lack of a perfect solution to facilitating the labeling process. Therefore, in this paper we present the development of our respiratory sound labeling software for normal breath sound and adventitious sound labeling, and this software could help for training deep learning-based models that require huge amounts of labeled acoustic data.

## 2. METHODOLOGY

### 2.1 Development Environment

In this study, we propose a respiratory sound labeling software to label wheezes, stridor, rhonchi, crackles, inspiration, expiration, noise, and normal sounds. The respiratory sound labeling software is developed based on Python 3.7 and PyQt5 using Visual Studio Code on a notebook computer, ASUSPRO P1440UF (ASUSTeK Computer Inc., Taipei, Taiwan). The configuration file is written in JSON format (.json). The users can change the preset parameters using a compatible text editor.

### 2.2 Labeling Software Comparison

We introduce two common sound labeling tools: MATLAB Audio Labeler and RX7. Then we compare our sound labeler to these two software packages. With MATLAB Audio Labeler (Figure 1 (a)), users can easily pre-define a label and perform one-click labeling and it supports multiple tracks of labels. However, it does not provide any spectrograms, and there is a perceptible latency time when playing audio. On the other hand, RX7 (Figure 1 (b)) provides spectrograms and corresponding parameter settings. It also has a dynamic display. Contrary to MATLAB Audio Labeler, it does not support custom predefined labels. Thus, users need to rename the label with default name and labeling becomes less flexible. In some cases, different labels overlap each other, and the range of each label is not clear. This makes labeling more difficult.

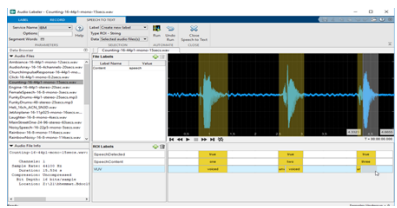  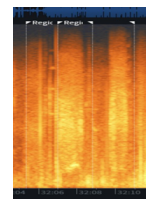

(a). MATLAB audio labeler　　　　　　　　　　　　(b). RX7

Figure 1. Labeling software comparison

**2.3 Features of Our Respiratory Sound Labeling Software**

Our labeling software integrates the merits of MATLAB Audio Labeler, and RX7. The Respiratory sound labeling software has following predefined sounds: normal, inspiration, expiration, wheeze, stridor, rhonchus, discontinuous, NBC (normal breathing circle), continuous and noise. The labeling software has following features:

1. Custom configuration:

The configuration file of the software is in JSON format. Users can modify the parameters in it to adjust the user interface (UI) or the requirements for a specific labeling task. For example, users can modify the parameters of short time Fourier transform and cutoff sound volume (dB) for visual display adjustment directly on the UI, and the display of the spectrogram will change accordingly, which is essential to make crackle sounds more clear (with better temporal solution by sacrificing spectral resolution) or to make continuous adventitious sounds more clear (with better spectral solution by sacrificing temporal resolution).

2. User-friendly interface:

On the UI, the software supports: forward and backward functions, drag and drop function, one-click labeling, spectrograms and multiple tracks with several different labels. The forward and backward functions allow users to review and label an exceptionally long audio file more quickly. The drag and drop function loads a series of the audio files in a more intuitive way. With our one-click labeling feature, it accelerates the labeling process. Spectrograms and multiple tracks can clearly display different labels. The software supports several different labels, which can be set in the configuration file.

3 Automatic saves:

The UI automatically saves the run-time labels so that if anything accidentally ceases or closes the software, the unfinished work can be recovered. When opening the audio files, the UI automatically reloads the corresponding historical labeling files. Additionally, users can stop labeling and close the UI at any time. When labelers go back to labeling, the UI then reads the auto-saved labels and lets the labelers start from the unfinished work left from last time. It eases the work of labeling an audio file that may last for hours.

4. File organization:

With our user identity management, it is easier to link users to the files they labeled. There is also an unlimited file list that shows which audio file is getting labeled. File switching is easier with the file list. Moreover, users click and store the audio currently under labeling to a database as gold standard sound. This is useful for reviewing the standard sound at any time whenever users get confused while labeling.

**2.4 Respiratory Sound Labeling Software Design and User Interface**

The fidelity of the labels of the inhalation, exhalation, and various adventitious sounds plays an important role in developing an accurate lung sound analysis algorithm. Nevertheless, labeling requires high labor effort and it is time-consuming. We made specific efforts to design the labeling software and its user interface in order to facilitate the labeling process and minimize the probability of labeling errors (Figure 2).

Respiratory sound labeling software is an annotation system with configurable user interface for respiratory sound specific needs. The respiratory sound labeling software enables users to interactively define and visualize ground-truth labels for respiratory sound data. It can be used to annotate audio by using preset labels and one-click labeling, and available for all common desktop operating systems. The software also supports multitrack labeling for several different sound status at the same moment. Users could freely add labels at any time position. They can select the audio files for annotation from the File List panel and then interactively label a set of audio files. These convenient features of the software allow research teams or users to decrease the time on creating respiratory sound labeling data for training a deep learning-based respiratory sound analysis model.

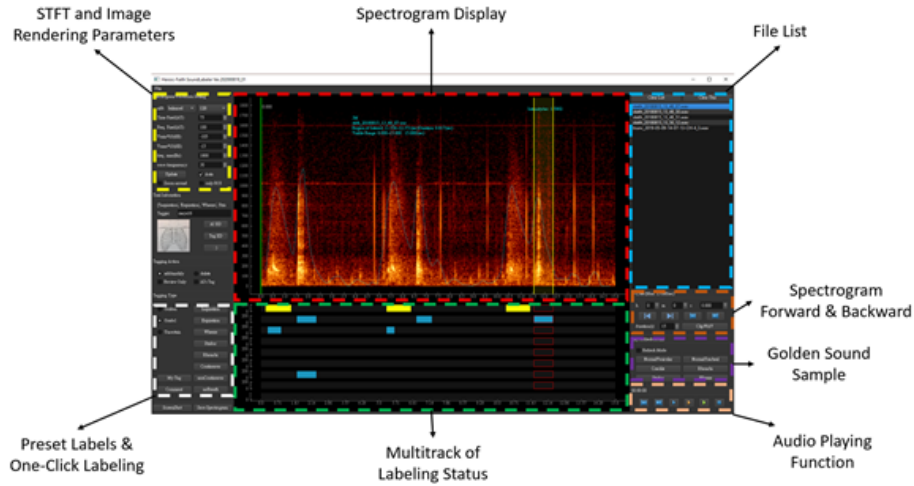

Figure 2. Respiratory sound labeling software interface

## 3. RESULTS

### 3.1 Respiratory Segments for Training and Evaluation

The lung sound data we collected were sampled at 4 kHz, and the bit depth was 16 bits. The audio files were recorded in WAVE (.wav) format. Since Pasterkamp, et al. [9] demonstrated that at least 15 seconds of continuous recording are required for verification and validation, we truncated the collected sound recordings to 15 seconds for subsequent analysis. Please refer to Table 1 for the summary of the number of the truncated 15-second recordings and the total duration.

For training a deep learning-based respiratory sound analysis model, we first need to generate a labelled data set. To aid the manual labelling, we implemented our proposed respiratory sound labeling software (Figure 1). As of October, 2019, we have labeled 9,765 15-second-long audio files of breathing lung sound, including 34,095 inhalation labels, 18,349 exhalation labels, 13,883 continuous adventitious sound labels (composed of 8,457 wheeze labels, 686 stridor labels and 4,740 rhonchus labels) and 15,606 discontinuous adventitious sound (all are crackles) labels. The statistics of the recordings and labels are listed in Table 1.

Table 1. Statistics of recordings and labels

|  | Number | Total duration (min) | Mean duration (sec) |
| --- | --- | --- | --- |
| Subjects | 279~517 |  |  |
| 15-sec recordings | 9,765 | 2441.25 |  |
| inhalation | 34,095 | 528.14 | 0.93 |
| exhalation | 18,349 | 292.85 | 0.96 |
| continuous adventitious sound | 13,883 | 191.16 | 0.83 |
| wheeze* | 8,457 | 119.73 | 0.85 |
| stridor* | 686 | 9.46 | 0.83 |
| rhonchus* | 4,740 | 61.98 | 0.78 |
| discontinuous adventitious sound | 15,606 | 230.87 | 0.89 |

*Wheezes, stridor, and rhonchi were combined to form continuous adventitious sounds.

### 3.2 Deep Learning Models and Experimental Results

In this study, we presented the benchmark results of convolutional bidirectional gated recurrent unit networks (CNN-BiGRU) and bidirectional gated recurrent unit networks (BiGRU) of inhalation, exhalation, and adventitious sounds detection.

The values of F1-score at the threshold where renders the best segment detection accuracy of each model on all tasks are tabulated in Table 2. According to Table 2, the CNN-BiGRU has a F1-score of 79.9% on inhalation segment detection and 86.0% on inhalation event detection. Best CASs detection can be reached based on the CNN-BiGRU with a F1-score of 53.3% on segment detection and 51.6% on event detection. The CNN-BiGRU has the best F1-score of 70.6% on DASs segment detection and the BiGRU has the best F1-score of 71.4% on DASs event detection.

Table 2. The F1-score of the segment detection and event detection for inhalation, CASs, and DASs detection.

| Models | n of trainable parameters | Inhalation | | CASs | | DASs | |
|---|---|---|---|---|---|---|---|
| | | F1-score | | F1-score | | F1-score | |
| | | Segment Detection | Event Detection | Segment Detection | Event Detection | Segment Detection | Event Detection |
| BiGRU | 552,769 | 79.8% | 85.7% | 26.9% | 25.6% | 70.3% | **71.4%*** |
| CNN-BiGRU | 5,240,513 | **79.9%*** | **86.0%*** | **53.3%*** | **51.6%*** | **70.6%*** | 70.0% |

* indicate the best F1-score among all the baseline models.

## 4. CONCLUSION

The goal of our work was to develop a software toolkit to help understand better the characteristics of different lungs sounds, ranging from normal sounds to pathological sounds, and perform labeling.

We achieved the following goals:

1. We developed a respiratory sound labeling software for analyzing and labeling lung sounds.
2. We made our database including 9,765 15-second-long audio files of lung sound, 34,095 inhalation labels, 18,349 exhalation labels, 13,883 CASs labels, and 15,606 DASs labels.
3. We developed an algorithm, CNN-BiGRU, for detecting, classifying, and characterizing a variety of physiologically relevant respiratory events, and evaluated our analysis techniques compared with BiGRU.
4. We reported that the CNN-BiGRU has the F1-scores of 79.9% on inhalation segment detection and 86.0% on inhalation event detection. CASs detection can be achieved based on the CNN-BiGRU model with a F1-score of 53.3% on segment detection and 51.6% on event detection. The CNN-BiGRU has the best F1-score of 70.6% on DASs segment detection and the BiGRU has a better F1-score of 71.4% on DASs event detection.

In conclusion, our results show that our proposed respiratory sound labeling software is capable of ground-truth annotation for audio, segmentation, and classification. A usability experiment shows that users could easily pre-define a label, perform one-click labeling, and overall facilitate the process of accurately labeling with our proposed respiratory sound labeling software.

## ACKNOWLEDGMENTS

This work is partly funded by Raising Children Medical Foundation, Taiwan. The author would like to acknowledge the National Center for High-Performance Computing (TWCC) in providing huge computing resources to facilitate this research. They also thank the All Vista Healthcare Center, Ministry of Science and Technology, Taiwan for the support.